\documentclass{article}

\usepackage{spconf,amsmath,graphicx}
\usepackage{amsmath,amsfonts,amssymb}
\usepackage{color}
\usepackage[french]{babel}
\usepackage[absolute]{textpos}
\AddThinSpaceBeforeFootnotes
\FrenchFootnotes

\title{Using hydrodynamical simulations of stellar atmospheres for periodogram standardization : Application to exoplanet detection}
\name{S. Sulis, D. Mary, L. Bigot}
\address{Laboratoire Lagrange, UMR CNRS 7293, Universit\'e C\^ote d'Azur, \\     
      Observatoire de la C\^ote d'Azur,  CS 34229, 06304 Nice, France}

\begin{document}
\ninept

\begin{textblock}{15.5}(0.2,0.2)
\noindent\small Published in the IEEE 2016 International Conference on Acoustics, Speech, and Signal Processing (ICASSP 2016), scheduled for 20-15 March 2016 in Shanghai, China. Personal use of this material is permitted. However, permission to reprint/republish this material for advertising or promotional purposes or for creating new collective works for resale or redistribution to servers or lists, or to reuse any copyrighted component of this work in other works, must be obtained from the IEEE. Contact: Manager, Copyrights and Permissions / IEEE Service Center / 445 Hoes Lane / P.O. Box 1331 / Piscataway, NJ 08855-1331, USA. Tel: + Intl. 908-562-3966.
\end{textblock}

\maketitle

\begin{abstract}
Our aim is to devise a detection method for exoplanet signatures (multiple sinusoids) that is both powerful and robust to partially unknown statistics under the null hypothesis. In the considered application, the noise is mostly created by the stellar atmosphere, with statistics depending on the complicated interplay of several parameters. Recent progresses in hydrodynamic (HD) simulations show however that realistic stellar noise realizations can be numerically produced off-line by astrophysicists. We propose a detection method that is calibrated by HD simulations and analyze its performances.
A comparison of the theoretical results with simulations on synthetic and real data shows that the proposed method is powerful and robust. 
\end{abstract}



 \vspace{-4mm}
\section{Introduction}
 \label{sec1}
  \vspace{-2mm}
This study is motivated by the long-standing challenge of detecting rocky low mass exoplanets. 
In this aim, future instruments with extremely low detector noise are being developed, giving access to massive time series of observations  (high sampling rates of typically five samples per minute, for several months to years, see \textit{e.g.} \cite{Ricker_2014,Rauer_2014,Pepe_2010}). 
We  focus on the so-called radial velocity data, where the planet signature   (if present) appears as one or a few sinusoidal components of weak amplitudes w.r.t. the noise level  \cite{Fischer_2014}. 
For the considered new observing facilities, the main noise source will not come from the instrument but from the stochastic activity of the star itself (convection, magnetic dynamo, stellar spots  and  oscillations). 

Sinusoid detection is another long-standing problem. In various fields, including indeed Statistics and Astronomy, a particularly rich amount of methods exists, a central piece of which is
the  (Schuster's) periodogram \cite{Schuster_1898}:
	\vspace{-2mm}
\begin{equation} \footnotesize
\label{eq1}
	P(\nu)   :=  \frac{1}{N} \Big| \sum_{j=1}^{N} X(t_j)\mathrm{e}^{-i2\pi\nu t_j} \Big|  ^2.
		\vspace{-2mm}
\end{equation}
In Eq.\eqref{eq1}, $X(t)$ will be  an evenly sampled time series obtained at $N$ epochs $t_j$, with sampling rate $\Delta t = t_{j+1}-t_j$, $\forall j \in \{1,\hdots,N-1\}$. \\
The type I errors (or false alarms, FA) of any test based on the ordinates of $P$  depend on the  statistics of these ordinates under the null hypothesis.  In practice, the statistics of the noise are often not (or only partially) known, and so are those of the $P$ ordinates. In this case, it is  difficult to assess how reliable is any claimed detection. 
%

In Statistical signal processing, some classical tests (\textit{e.g.} \cite{Fisher_1929,Chiu_1989,Shimshoni_1971}) partially cope with this issue by guaranteeing the control of the FA rate whatever the unknown noise variance, provided that the noise is  Gaussian,  independent and identically distributed (i.i.d.).
When the noise is not white but colored in some unknown manner, a classical approach consists in calibrating $P(\nu)$ by some estimate $\widehat{P}(\nu)$, leading to a frequency-wise standardized periodogram of the form  $\frac{P(\nu)}{\widehat{P}(\nu)}$. Usually, the noise Power Spectral Density (PSD)  has to be estimated from the data. This can be done through non-parametric methods (\textit{e.g.} local SNR \cite{Zheng_2012}, modified periodogram smoothers  \cite{Hannan_1961,Priestley_1981}, robust M-estimators  \cite{Sachs_1993,Sachs_1994}) or parametric methods (\textit{e.g.}  iterative Yule-Walker  \cite{Chiu_1990}, ratio of autoregressive (AR) spectral estimates \cite{White_1999}, balanced model truncation \cite{Gryca_1998}). Even if some of these estimators are asymptotically unbiased, the unavoidable injection of  estimation noise in the denominator of the standardized periodogram makes  the statistical characterization of the test statistics difficult. One can resort to Monte Carlo  or  bootstrap simulations \cite{Zoubir_1993}  to evaluate the  thresholds empirically (see  \cite{Li_2014}, Chap.7 for examples of gaps between theoretical and empirical thresholds), but this procedure may not be tractable for massive time series (large $N$). \\
In Astronomy, preprocessing stages aimed at ``whitening'' the noise  (\textit{e.g.} with  local trend filtering or line removal with the CLEAN method \cite{Hogbom_1974}) or based on various ARMA (AR-Moving Average) noise models  \cite{Tuomi_2012} are most often applied to the data  before conducting the detection test. Such procedures pose the same question of robustness about the actual FA rate at which the test is conducted.  In presence of unknown colored noise, the reliability of a claimed detection of a low mass telluric planet  is thus difficult to assess (see for instance the recent and controversial case of $\alpha$ Centauri B  \cite{Dumusque_2012,Hatzes_2013}). \\
\indent In the present work,  we do not attempt to build  dedicated parametric noise models. We choose instead to exploit recent progresses in HD simulations \cite{Bigot_2011}. These results  demonstrate that reliable time series of the stellar noise can be simulated by numerical codes, which account for the complex interplay of various astrophysical processes in the star's interior. We  assume that a training data set (in the form of time series) is available and use this information to standardize the periodogram. The data set is considered unbiased, but possibly limited in size, because  HD simulations are heavy.  In fact, simulating 100 days at high sampling rate requires around 3 calculation months over 120 CPU. Consequently, realistic numbers of simulated light curves available in practice will not be more than, say,  a hundred. This raises the question of the impact of estimation noise in the proposed standardization approach.\\
 Addressing this question first requires,  for the two hypotheses of our model (Sec.\ref{sec2}), the investigation of the  statistics of the classical, averaged  and standardized periodograms. This is the purpose of  Sec.\ref{sec3}, where the use of asymptotic results is motivated by the large duration and high sampling rate (large $N$) considered here.  The second step is to select several tests (Sec.\ref{sec4}) for which the benefits gained from the proposed standardization can be highlighted and quantified. We  opt for a sample of classical tests covering the different cases of  single and multiple sinusoids detection. At this stage we are in position to derive  the tests statistics 
 and the corresponding FA (Sec.\ref{sec5}) and detection (Sec.\ref{sec6})  rates. The last step is a numerical evaluation of the theoretical results
 , the method performances and its actual robustness (Sec.\ref{sec7}). 
 \vspace{-3mm}
\section{Statistical model}
 \label{sec2}
\vspace{-2mm} 
We consider two hypotheses :
\begin{equation} \footnotesize
  \left\{
      \begin{aligned}
	 \text{ ${\cal{H}}_0$ : } X(t_j) &= \epsilon(t_j) \\
	 \text{ ${\cal{H}}_1$ : } X(t_j) &=  \sum_{i = 1}^{N_s} \alpha_i \sin(2\pi f_i t_j+\phi_i)+ \epsilon(t_j) \\
      \end{aligned}
    \right.
    \label{hyp}
    	\vspace{-2mm}
\end{equation}
where $X(t_j=j\Delta t)$ is the evenly sampled data time series 
and $\epsilon(t_j)$ a zero-mean second-order stationary Gaussian noise with  PSD $S_\epsilon(\nu)$ and autocorrelation function $r_\epsilon$, for which  $\inf ( S_\epsilon(\nu) ) > 0$ and  $\sum_k | r_\epsilon(k) | < \infty$. The $N_s$  amplitudes $\alpha_i$, frequencies  $f_i$ and phases $\phi_i$ represent the planet signatures and are unknown. 
\vspace{-2mm}
%
\section{Periodograms' statistics : asymptotics}
 \label{sec3}
  \vspace{-2mm}
\subsection{Classical (Schuster's) periodogram}
  \vspace{-2mm}
  Without loss of generality and to simplify the presentation, $N$ is even and the considered frequencies in Eq.\eqref{eq1} belong to the subset of $\frac{N}{2}-1$  Fourier frequencies $\nu_k=\{\frac{k}{T}\}$
where $k \in \Omega := \{ 1,\hdots,  {\frac{N}{2}-1}\}$.
Asymptotically, the  periodogram $P$ is an unbiased but inconsistent estimate of the PSD \cite{Brillinger_1981}. 
    However, under the above assumptions on $\epsilon$, the periodogram ordinates at different frequencies $\nu_k$ and $\nu_{k'}$ are asymptotically independent  \cite{Li_2014}. \\
Under  ${\cal{H}}_0$, the asymptotic distribution of $P$ is (\cite{Brillinger_1981}, theorem 5.2.6) :
\begin{equation}  \small
P(\nu_k |{\cal{H}}_0) \sim \frac{S_\epsilon(\nu_k)}{2} \chi_2^2 ,~~~~~~~~~~ \forall k ~ \in ~ \Omega.
    \label{dist}
\end{equation}
The $\chi^2_2$ is a $\chi^2_1$ at $k=0,\frac{N}{2}$. We restrict to $\Omega$ for simplicity. \\
Under ${\cal{H}}_1$, the distribution of $P(\nu_k)$ can be found in the complex case in \cite{Li_2014}, Corollary 6.2.
Using this corollary and Euler's formula for the sines in model (\ref{hyp}), we obtain that, for large $N$,  $P(\nu_k)$ is approximately distributed as:
\begin{equation} \small
		P(\nu_k | {\cal{H}}_1) \sim \frac{S_\epsilon(\nu_k) }{2} \chi_{2,  \lambda_k}^2 ,~~~~~~~~~~ \forall k ~ \in ~ \Omega, 
    \label{dist2}
\end{equation}
with $\lambda_k = \lambda(\nu_k)$
a non-centrality parameter expressing the spectral leakage of all signal frequencies at location $\nu_k$. Generally, this leakage is caused by the fact that the $\{f_i\}$ do not belong  to the natural Fourier grid defined by $\Omega$. Denoting by  $K_N(\nu)=\Big( \frac{\sin(N \pi \nu)}{N\sin(\pi \nu)} \Big)^2$ the Fej\'er  kernel (or spectral window), the asymptotic expression of the parameters 
$\lambda_k $, $k\in \Omega$ for model (\ref{hyp}) writes:
\begin{equation} 
  \label{lambda}  \footnotesize
	  \begin{aligned} 
             &\lambda(\nu_k)= \frac{N}{2S_\epsilon(\nu_k)} \sum_{i=1}^{N_s} \alpha_i^2 \Big[ K_N(f_i - \nu_k) + K_N(f_i + \nu_k)  \\
             &+2 \frac{\sin(N \pi  (f_i-\nu_k))}{N\sin(\pi  (f_i-\nu_k))}  \frac{\sin(N \pi  (f_i+\nu_k))}{N\sin(\pi  (f_i+\nu_k))} \cos(2\pi(N+1)f_i + 2 \phi_i) \Big].
               \end{aligned}
\end{equation}	  
\vspace{-0.8cm}
\subsection{Averaged periodogram}
We assume that a training data set $\mathcal{T}$ of independent realizations of the stellar noise is available. This set  is obtained by HD simulations and composed with $L$ times series $X_\ell$ sampled on the same grid as the observations : $\mathcal{T}=\{X_\ell(t_j), j=1,\hdots,N\}$, $\ell=1,\hdots,L$. \\
A straightforward, consistent and unbiased estimate of the noise PSD can be obtained by the averaged periodogram \cite{Bartlett_1950}:
$$ \small
	\overline{P}(\nu |  {\cal{H}}_0)  : =  \frac{1}{L} \sum_{\ell=1}^{L}  \frac{1}{N} \Big| \sum_{j=1}^{N} X_\ell(t_j)\mathrm{e}^{-i2\pi\nu t_j} \Big| ^2,
$$
whose asymptotic distribution can be easily obtained using \eqref{dist} as:
\begin{equation} \small
\overline{P}(\nu_k | {\cal{H}}_0) \sim  \frac{S_\epsilon(\nu_k)}{2L}  \chi_{2L}^2,  ~~~~\forall k ~ \in ~ \Omega.   
    \label{dPmoy}
\end{equation}
Note that in this setting, any source of bias (caused for instance by imperfect HD simulations) is left out of scope of this study.
The focus is on the stochastic estimation noise caused by the finiteness of $\mathcal{T}$, which is encapsulated in $L$ and impacts the distribution of $\overline{P}$.
  \vspace{-4mm}
\subsection{Standardized periodogram}

The standardized periodogram considered here is defined as:
\begin{equation} \small
\label{eq_pr}
	 \widetilde{P}(\nu_k) := 	\frac{P(\nu_k)}{\overline{P}(\nu_k)}.
\end{equation}
As the numerator and denominator are independent variables with known asymptotic distributions, assessing the distribution of their ratio is straightforward. 
The ratio of two  independent random variables (r.v.)  $V_1\!\sim\! \chi_{d_1}^2$ and $V_2\!\sim\! \chi_{d_2}^2$ follows a Fisher-Snedecor law  noted $F(d_1,d_2)$ with ($d_1,d_2$) degrees of freedom  : $\frac{V_1/d_1}{V_2/d_2}\!\! \sim\!\! F(d_1,d_2)$ \cite{Abramowitz_1972}.
Consequently, from (\ref{dist}) and (\ref{dPmoy}), the asymptotic distribution of $ \widetilde{P}$ under $\mathcal{H}_0$ is:
\begin{equation} \small
 \widetilde{P}(\nu_k | {\cal{H}}_0)\!\sim \frac{S_\epsilon(\nu_k)\chi_2^2/2}{S_\epsilon(\nu_k)\chi_{2L}^2/2L}  \sim F(2,2L),  ~~~~~ \forall k ~ \in ~ \Omega.
    \label{dist3}
\end{equation}
Similarly, under $\mathcal{H}_1$, we deduce the asymptotic distribution of $ \widetilde{P}$ with (\ref{dist2}) and  (\ref{dPmoy})  :
\vspace{-2mm}		
\begin{equation} \small
		\widetilde{P}(\nu_k  | {\cal{H}}_1)\sim \frac{\chi_{2, \lambda_k}^2/2}{\chi_{2L}^2/2L} \sim F_{\lambda_k}(2,2L),   ~~~ \forall k ~ \in ~ \Omega,
	\label{dist_H1}
\end{equation}
with $\lambda_k$ given by (\ref{lambda}).
These results call for two remarks. First, an $F$ distribution similar to that of  \eqref{dist3} was obtained in \cite{Lu_2005} when using ratios of the form $P(\nu_k)/P(\nu_l)$, $k\neq l$ (for symmetry testing purposes). Second, \eqref{dist3} shows that the distribution of the standardized periodogram is independent of the (partially unknown) noise PSD, which is indeed a necessary condition for  a robust detection test.
\vspace{-0.2cm}
\section{Statistical tests}
 \label{sec4}
  \vspace{-2mm}
 The first three tests below are considered for comparison purposes. The three other tests are the counterpart of classical tests applied to $\widetilde{P}$ instead of $P$ (many more such tests could be devised).\\
  In the following, the order statistics of the periodograms will be noted with parentheses. For instance, the order statistics of $P$ are:\\ 
 $ \displaystyle{\min_k} ~ P(\nu_k)=P_{(1)}<P_{(2)}<\hdots<P_{(\frac{N}{2}-1)} = \max_k P(\nu_k)$.
 \vspace{-2mm}
 \subsection{ Classical tests}
 \vspace{-2mm}
  \label{sec31}
 Perhaps the most classical reference test (including in Astronomy) is the Fisher test : 
\vspace{-0.2cm}
\begin{equation} \small
\label{eq3b}
{T}_{Fi}  \mathop{\gtrless}_{\mathcal{H}_0}^{\mathcal{H}_1} \gamma, \quad {\textrm{with}}\quad {T}_{Fi}:= \frac{P_{(\frac{N}{2}-1)}}{\displaystyle{\sum_{k ~ \in ~ \Omega} P_{(k)}}},
\vspace{-0.2cm}
\end{equation}
where $\gamma\in \mathbb{R^+}$ is a threshold that determines the FA rate.
 This test is robust to an unknown noise variance \cite{Schwarzenberg_1998}, but the noise must be white Gaussian.
 When this is the case, the denominator of (\ref{eq3b}) is an asymptotically unbiased estimate of the PSD (to a constant) \cite{Priestley_1981}. 
Under $\mathcal{H}_1$, for a model involving a single sinusoid $f_1$  on the Fourier grid, the Generalized Likelihood Ratio (GLR) test corresponds to $P_{(N/2-1)}\mathop{\gtrless}_{\mathcal{H}_0}^{\mathcal{H}_1}\gamma$ \cite{Kay_1998}.  So, under $\mathcal{H}_0$, the Fisher test can be seen as a standardization of the GLR by the estimated variance  $\hat{\sigma}^2 :\propto  {\sum_{k ~ \in ~ \Omega}P(\nu_k)} $. 

In the case of multiple sinusoids  ($N_s > 1$), the performances of the Fisher test are known to decrease, owing to the perturbations of sinusoids in the noise variance estimation. Many alternatives exist, \textit{e.g.} \cite{Siegel_1980,Priestley_1981}.  These tests attempt to better estimate the noise level by excluding a number $N_c$ of ordinates presumably contaminated by the sinusoids (see \cite{Priestley_1981,Li_2014,Brillinger_1981}). 
Two such tests, offering the same robustness against unknown variance, are the robust Fisher test \cite{Li_2014}  and the Chiu test \cite{Chiu_1989} defined by ${T}_{Fi,rob}\mathop{\gtrless}_{\mathcal{H}_0}^{\mathcal{H}_1} \gamma$ and $ {T}_{Ch}  \mathop{\gtrless}_{\mathcal{H}_0}^{\mathcal{H}_1} \gamma$, where : 
\vspace{-0.2cm}
\begin{equation} \small
\label{test_cl}  
 {T}_{Fi,rob}:=  b_r (\frac{N}{2} -1) ~ r  ~ {\frac{P_{(\frac{N}{2}-1)}}{\displaystyle{\sum_{k=1}^{\frac{N}{2}-1-N_c}P_{(k)}}}},  ~~~
 {T}_{Ch}:= \frac{P_{(\frac{N}{2}-N_c)}}{\displaystyle{\sum_{k=1}^{\frac{N}{2}-1-N_c}P_{(k)}}},
\end{equation} 
with $r =  \frac{\frac{N}{2} -1 - N_c}{\frac{N}{2} -1}$ and $\small b_r = 1 + r^{-1}(1-r) \log(1-r)$ . Note that the last two approaches pose the question of the choice of $N_c$.
Not fixing $N_c$ in advance but estimating this parameter from the data may lead to a more powerful test, but at the cost of a weaker control of the FA rate (as $N_c$ is random). To avoid this complication, the tests will be compared in the numerical study for $N_c$ set to $N_s$.\\
\vspace{-6mm}
\subsection{ Tests based on the standardized periodogram}
 \label{sec32}
\vspace{-2mm}
Under ${\cal{H}}_0$ and in the asymptotic regime, $\widetilde{P}$ in (\ref{eq_pr}) is i.i.d. since the ratio cancels out the frequency dependence on the PSD. \\ If $N_s = 1$, the following simple test is thus likely to be powerful:
\vspace{-2mm}
\begin{equation} \small
\label{eq3bis}
\widetilde{T} ~ \mathop{\gtrless}_{\mathcal{H}_0}^{\mathcal{H}_1} \gamma, \quad {\textrm{with}}\quad \widetilde{T}:= \widetilde{P}_{(\frac{N}{2}-1)}.
	\vspace{-2mm}
\end{equation}
Similarly, the discussion above suggests to  consider Fisher's approach and to apply  test  (\ref{eq3b}) to (\ref{eq_pr}):
\vspace{-3mm}
\begin{equation} \small
\label{eq3ter}
\widetilde{T}_{Fi}  \mathop{\gtrless}_{\mathcal{H}_0}^{\mathcal{H}_1} \gamma, \quad {\textrm{with}}\quad \widetilde{T}_{Fi}:= \frac{\widetilde{P}_{(\frac{N}{2}-1)}}{\displaystyle{\sum_{k \in \Omega} \widetilde{P}_{(k)}}}.
\vspace{-2mm}
\end{equation}
Finally, in the case of several sinusoids, the good results of Chiu's test suggest to look for deviations in the region of the $N_c^{\textrm{th}}$ largest $\widetilde{P}$ ordinate, $\widetilde{P}_{(\frac{N}{2}-N_c)}$,  instead of the largest one, $\widetilde{P}_{(\frac{N}{2}-1)}$ :
\vspace{-2mm}
\begin{equation} \small
\label{eq17}
\widetilde{T}_{N_c}  \mathop{\gtrless}_{\mathcal{H}_0}^{\mathcal{H}_1} \gamma, \quad {\textrm{with}}\quad \widetilde{T}_{N_c}:= \widetilde{P}_{(\frac{N}{2}-N_c)}.
\end{equation}
	\vspace{-8mm}
\section{Statistics under  ${\cal{H}}_0$ and false alarm rate}
 \label{sec5}
\vspace{-2mm}
The accurate control of the false alarm probability ($\rm{P_{FA}}$) in case of partially unknown colored noise is a critical point. We now show that
while this control is (not surprisingly) problematic with classical tests like (\ref{eq3b}, \ref{test_cl}),  the proposed  tests (\ref{eq3bis}, \ref{eq17}) allow such a control. \\
We investigate first the tests ${T}_{Fi}, \widetilde{T}_{Fi}$ and $\widetilde{T}$ (designed for single sinusoid detection).
Under $\mathcal{H}_0$, their test statistics involve the largest value of a set of  $\frac{N}{2}-1$ r.v., ${T}_{Fi}(n), \widetilde{T}_{Fi}(n)$ and $ \widetilde{T}(n)$, whose definitions and distributions are given by, using (\ref{dist2}) and (\ref{dist3}) :
\begin{equation} \footnotesize
	\begin{aligned}
	{T}_{Fi}(n) ~~&:= \frac{P(\nu_n)}{\displaystyle{\sum_{k ~ \in ~ \Omega}P_{(\nu_k)}}}  ~ \stackrel{as,ni.ni.d.}\sim \frac{\chi^2_2S_\epsilon(n)/2}{\displaystyle{\sum_{k ~ \in ~ \Omega}}\chi^2_2S_\epsilon(k)/2  },\\ 
	 \widetilde{T}_{Fi}(n) &:= \frac{\widetilde{P}(\nu_n)}{\displaystyle{\sum_{k ~ \in ~\Omega}\widetilde{P}{(\nu_k)}}}  \stackrel{as,ni.i.d.}{\sim} \frac{F(2,2L)}{  \displaystyle{\sum_{k ~ \in ~\Omega}} {F(2,2L) }}, \\
	 \widetilde{T}(n) ~~~& :=\;\;\;\;\;\;\widetilde{P}(n)\;\;\;\; \stackrel{as,i.i.d.}{\sim} ~~~~~~~ F(2,2L),	\end{aligned}
	\label{eq20}
\end{equation}
where ``\textit{as,ni.ni.d.}'' means asymptotically non independent and non identically distributed. 
In the first  case, each r.v. ${T}_{Fi}(n)$ clearly depends on the unknown PSD.  In the second case, each r.v. $ \widetilde{T}_{Fi}(n)$ is a ratio  of a $F$ variable over a sum of $F$ variables. To our knowledge, there is no analytic characterization of the resulting distribution. Hence, in both cases, assessing the distribution of the maximum of these variables is problematic.
In the third case, the $\widetilde{T}(n)$ r.v. are independent and follow a $F$ distribution with known density $\varphi_F$ \cite{Abramowitz_1972}. Using the incomplete Beta function $B(d_1,d_2)$  and noting that $B(1,L)=\int_0^1(1-t)^{L-1}\textrm{d}t=\frac{1}{L}$ we obtain:
$$\footnotesize
\varphi_F(\gamma,2,2L)=\frac{1}{B(1,L)}\cdot\frac{1}{L}\cdot \Big(1+\frac{\gamma}{L}\Big)^{-L-1}=\Big(1+\frac{\gamma}{L}\Big)^{-L-1}.
$$
It can be checked that the norm of $\varphi_F(\gamma,2,2L)$  is $1$. 
Turning to the CDF  $\Phi_F(\gamma,2,2L)$ of $\widetilde{T}(n)$,
we obtain by using $F(\gamma,d_1,d_2) \!= \!I_{ \frac{d_1 \gamma}{d_1 \gamma + d_2}} (\frac{d_1}{2}, \frac{d_2}{2})$ \cite{Abramowitz_1972} or by direct integration of $\varphi_F(\gamma,2,2L)$ :
 \vspace{-2mm}
\begin{equation}
 \small
\Phi_F(\gamma,2,2L)=1-\Big(\frac{L}{\gamma+L}\Big)^L.
	\vspace{-2mm}
\label{laphi}
\end{equation}
The $\rm{P_{FA}}$ can be computed thanks to the  asymptotic independence of the $\{\widetilde{T}(n)\}$:
\begin{equation} \small
\begin{aligned} 
	\!\rm{P_{FA}}\!(\widetilde{T},\gamma) &:=
	\textrm{Pr}\; (\max_n \widetilde{T}(n)\! > \!\gamma | {\cal{H}}_0\!)=1 -\displaystyle{ \prod_n} \textrm {Pr}\, (\widetilde{T}(n) \leq \gamma | {\cal{H}}_0\!)\!\! \\
			    &\!\!\!\!\!\!\!\!\!\!\!\!\!\!\!\!\!\!\!\!\!\!\!\!\!\!
			   = 1 - \Big(\Phi_F(\gamma,2,2L)\Big)^{\frac{N}{2}-1} 
			     = 1-  \Big( 1-\Big(\frac{L}{\gamma+L}\Big)^L\Big)^{\frac{N}{2}-1},
\end{aligned}
\label{pfa}
 \end{equation}
which is  a remarkably simple  expression. \\
We turn now to the tests   $T_{Fi,rob},  T_{Ch}$ and  $\widetilde{T}_{N_c}$ (designed for multiple sinusoids detection). 
The  asymptotic distributions of the first two tests are  known under the WGN assumption \cite{Chiu_1989}. Using the same reasoning
as above for $T_{Fi}$, it is easy to see that the noise PSD  affects the distribution of the order statistics involved in these tests, with uncontrollable impact on the $\rm{P_{FA}}$.
Turning to $\widetilde{T}_{N_c}$, remark that the number $K$ of ordinates $\widetilde{P}$ larger than $\gamma$ follows a binomial distribution:
 $K\sim \mathcal{B}(\frac{N}{2}-1,1-\Phi_{F}(\gamma,2,2L))$, from which, with \eqref{laphi}, we obtain :  
 \vspace{-3mm}
\begin{equation}\small  
\vspace{-5mm}
\begin{aligned} 
	&\!\rm{P_{FA}}\!(\widetilde{T}_{N_c},\gamma) \!:=\!
	\textrm{Pr}\; (\! \widetilde{T}_{N_c}\! > \!\gamma | {\cal{H}}_0\!)\! = 1- \sum_{i=0}^{N_c-1} \textrm{Pr}\; ( K=i) \\
	&= 1- \sum_{i=0}^{N_c-1} {N/2-1 \choose i}  \Big( \frac{L}{\gamma+L}\Big)^{Li} \Big(1-(\frac{L}{\gamma+L})^L\Big)^{\frac{N}{2}-1-i}.
\end{aligned}
\label{pfa_Tor}
 \end{equation}
\section{Statistics under  ${\cal{H}}_1$  and detection  rate}
 \label{sec6}
\vspace{-2mm}
Under  ${\cal{H}}_1$,  the same approach shows that one can not control the distribution of the r.v. ${T}_{Fi}(n)$ and $\widetilde{T}_{Fi}(n)$ and consequently the associated detection probabilities  ($\rm{P_{DET}}$). In contrast, from (\ref{dist_H1}) and using the same reasoning  as for the 
 $\rm{P_{FA}}$,  the $\rm{P_{DET}}$ of $\widetilde{T}$ is:
  \begin{equation} \small
	\label{Eq_Pdet}
	\rm{P_{DET}} (\widetilde{T}, \gamma) := \textrm{Pr} (\max_n \widetilde{P}(n)  > \gamma | {\cal{H}}_1) = 1 - \displaystyle{ \prod_{k \in \Omega}} \Phi_{F_{\lambda_k}}(\gamma, 2,2L).
\end{equation}	
\vspace{-2mm}
With (\ref{pfa})  the relationship $\gamma(\rm{P_{FA}})$ for $\widetilde{T}$ can be derived as:
 \begin{equation} \footnotesize
\label{Eq_gam}
		\gamma(\widetilde{T},\rm{P_{FA}}) =  L  \Big[ \Big( 1 - ( 1 - \rm{P_{FA}})^{\frac{1}{\eta}} \Big)^{-\frac{1}{L}} -1 \Big],   
\end{equation}	
{where $\eta = \frac{N}{2}-1$}. With (\ref{Eq_Pdet}) and (\ref{Eq_gam}), we  deduce :
 \begin{equation} \footnotesize
\label{Eq20}
	\rm{P_{DET}}(\widetilde{T},\rm{P_{FA}}) =  1 - \displaystyle{ \prod_{k\in \Omega }}  \Phi_{F_{\lambda_k}}(L  [ ( 1 - ( 1 - \rm{P_{FA}})^{\frac{1}{\eta}} )^{-\frac{1}{L}} -1 ], 2,2L),
\end{equation}	
 which can be used to compute the ROC curves.\\
 The function $\gamma \mapsto \rm{P_{DET}} (\gamma)$ of the test $\widetilde{T}_{N_c}$  can be deduced similarly to (\ref{pfa_Tor}), with the difference that $K$ is no longer binomial owing to the $\lambda_k$. Denote by ${\Omega}^{(i)}$ one particular combination of $i$ indices taken in $\Omega$ and
  $\overline{\Omega}^{(i)}:=\Omega \backslash \Omega^{(i)}$ the set of remaining indices. Let $\{\Omega^{(i)}_1,\hdots , \Omega^{(i)}_i\}$ (resp. $\{\overline{\Omega}^{(i)}_1,\hdots, \overline{\Omega}^{(i)}_{\frac{N}{2}-1-i}\}$) denote the indices in  two such combinations, and let   $\Omega^i$ (resp. $\overline{\Omega}^{i}$) be the set of all the  $\{{\Omega}^{(i)}\}$   (resp. of all $\{\overline{\Omega}^{(i)}\}$). With these notations we obtain:
 \begin{equation} \footnotesize 
 		\rm{P_{DET}}(\widetilde{T}_{N_c},\!\gamma) \!=\! 1 -\!\! \displaystyle{ \sum_{i=0}^{N_c - 1} \!\!\sum_{\Omega^{(i)} } \!\prod_{k=1}^{ i  } } \!\!\Big(\!1\! - \Phi_{F_{\lambda_{\Omega^{(i)}_k}}\!\!^{\!\!\!\!\!\!\!\!\!(\gamma, 2,2L)}} \!\!\Big) \displaystyle{ \!\!\! \prod_{k'=1}^{ \frac{N}{2}-1-i }} \Phi_{F_{\lambda_{\overline{\Omega}^{(i)}_{k'}}}\!\!^{\!\!\!\!\!\!\!\!\!(\gamma, 2,2L)}},
		\label{lagrosse}
\end{equation}	
which is typographically heavy but can be used to compute ROC curves (in Eq.(\ref{lagrosse})  the non-centrality parameters are given by (\ref{lambda})).
 \vspace{-4mm}
\section{Numerical simulations}
 \label{sec7}
\begin{figure}[htb!]   
	\hspace{-4mm}\includegraphics[width= 9cm,height=75mm]{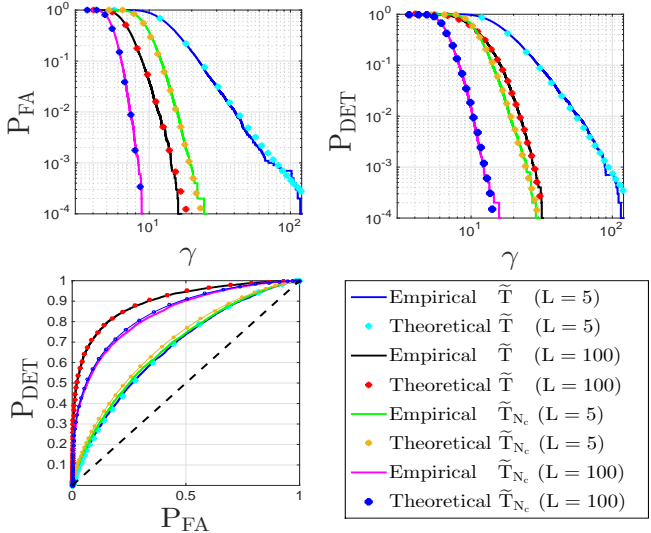}	
		\vspace{-9mm}
	\caption{\small Validation of the theoretical results for $\widetilde{T}$ and $\widetilde{T}_{N_c}$ by MC simulations. Parameters : $N = 1024$,  $\Delta t = 60$ s, $N_s = N_c = 3$, $ L=[5,100]$,  $\alpha_i = 0.1$ m/s for all sines, $f_i = [5, 5.75, 6.50]$ mHz.}
	\label{Fig1}
		\vspace{-3mm}
\end{figure}

\begin{figure}[htb!]
\vspace{-2mm}
	\hspace{-.4cm}\includegraphics[width=97mm,height=60mm]{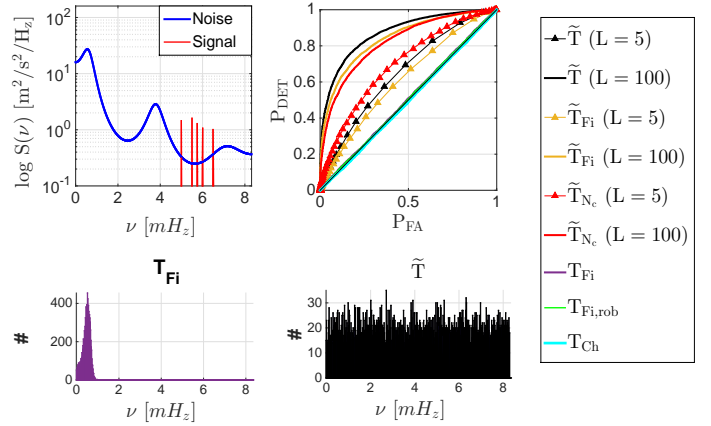}
	\vspace{-8mm}
	\caption{\small (a)  PSD of  AR noise (blue) and sines (red). (b)    ROC curves in case of signal frequencies falling into the PSD ``valley'' region : $ N_c = N_s = 5$,  $\alpha_i = 0.08$ m/s for all sines (the apparent amplitude difference is caused by the different leakage affecting these sines), $f_i = [5, 5.5, 5.75, 6, 6.5]$ mHz, with $10^4$ simulations, $N = 1024$,  $\Delta t = 60$ s, $ L= [5,100]$. (c, d) Histograms of the frequency distribution under ${\cal{H}}_0$ of $T_{Fi}=\displaystyle{\max_n}~  T_{Fi}(n)$ and $\widetilde{T} =  \displaystyle{\max_n}~ \widetilde{T}(n)$.  }
	\label{Fig2}
	\vspace{-8mm} 
\end{figure}

\begin{figure}[htb!]
	\hspace{-.4cm}\includegraphics[width=95mm,height=45mm]{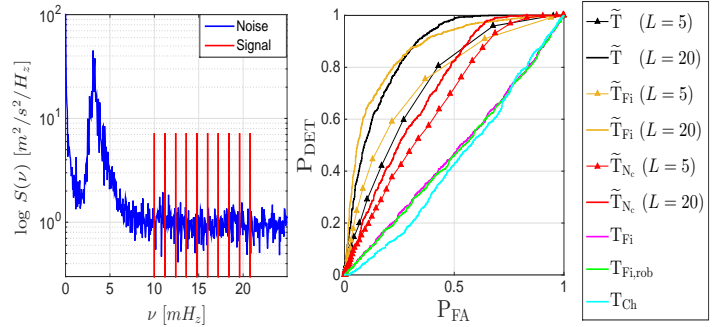}	
		\vspace{-8mm}
	\caption{\small Test on GOLF data. Parameters : $N = 10^3$,  $4452$ simulations for $L = 5$ and $1272$ simulations for $L = 20$, $\Delta t = 20$ s, $N_s = N_c = 10$, $\alpha_i = 0.17$ m/s, $\{f_i\}$ equally spaced in $[10; 20]$ mHz. }
	\label{Fig3}	
\end{figure}

  \vspace{-2mm}
  For the purpose of making empirical ROC curves, we first consider a noise corresponding to an autoregressive process : AR(6). 
The shape of its PSD is representative of a stellar PSD, with local variations (stellar oscillations) and  higher energy at low frequencies (convection, magnetic activity).   
 Under  ${\cal{H}}_1$, we consider the most general case $N_S > 1$. 
First of all, we evaluate the reliability of the $\rm{P_{FA}}(\gamma)$, $\rm{P_{DET}}(\gamma)$ and $\rm{P_{DET}}(\rm{P_{FA}})$ expressions for the $\widetilde{T}$ and $\widetilde{T}_{N_c}$ tests (Fig.\ref{Fig1}). We compare here our theoretical results (\ref{lambda}), (\ref{pfa}), (\ref{pfa_Tor}), (\ref{Eq_Pdet}), (\ref{Eq20}) and (\ref{lagrosse}) with $10^4$ Monte Carlo (MC) simulations. The theoretical expressions are in agreement with the MC simulations and the test performances logically increase with $L$ (as the estimation noise decreases). 
	\vspace{-4mm}
\\In a second experiment, we place the different signal frequencies in a ``valley''  of the noise PSD (Fig.\ref{Fig2}.a). We calculate the empirical ROC curves (Fig.\ref{Fig2}.b) for  the  tests under study (see (\ref{eq3b}) to (\ref{eq17})).
The figure shows that the performances of tests  using  $\widetilde{P}$ are better than those based on $P$ in this configuration (violet, green and  cyan curves on the diagonal). 
To gain more insight on the relative tests performances, we compare the frequency distribution under ${\cal{H}}_0$ of $T_{Fi}$  and $\widetilde{T}$ (Fig.\ref{Fig2}.c, d).
For $T_{Fi}$, the FA repartition is not uniform and increases in the PSD regions of larger energy ($\nu <1$ mH$_z$).
 When signals frequencies happen to fall into these zones, tests based on $P$ are favored, but when they fall outside such regions their power vanish.  In contrast, the  $\widetilde{T}$ test allows a good detectability over all the frequency range, with performances close to the asymptotic one ($L\rightarrow \infty$, no estimation noise) for $L\approx 10^2$.
In brief, the tests $\widetilde{T}$ and $\widetilde{T}_{N_c}$ allow to control precisely the $\rm{P_{FA}}$, they have good power and their  performances increase with $L$.   
\\ Finally, we apply the tests (\ref{eq3bis}) and (\ref{eq17})  to real solar data \cite{Garcia_2005} (Fig.\ref{Fig3}). These data have been collected for 18 years with an even sampling rate.  As this data set is the largest one currently available we use it to compute ROC curves on shorter time series of $N=1000$ samples extracted from the data set.  In order to be as close as possible to the considered setting (perfect HD simulations of the stellar noise), we use part of the data to standardize the periodogram in (\ref{eq1}). The left panel of Fig.\ref{Fig3} superimposes the solar PSD (blue) with the introduced signals (red). The right panel displays the ROCs curves. 
The proposed standardized tests $\widetilde{T}$,  $\widetilde{T}_{Fi}$ and $\widetilde{T}_{N_c}$ (black, yellow and red) are more powerfull  than the others. 
        \vspace{-4mm}
   \vspace{-6mm}
\section{Conclusion}
 \vspace{-2mm}
   We have investigated a detection method based  on periodogram standardization through HD-simulation to counteract the impact of  colored noise. We have  analyzed its statistical performances and  highlighted the shortcomings of tests ignoring the frequency dependence of the noise PSD. In contrast, the proposed standardization  leads to a robust detection method in the sense that  the $\rm{P_{FA}}$ is reliable and independent of the noise PSD. All theoretical results have been obtained in the asymptotic regime, but they appears to be a good approximation for relatively low values of $N$.
    \\
  \footnotesize \\
{\bf Acknowledgement} 
\textit{ We are grateful to Thales Alenia Space,  PACA region and CNRS project DETECTION/Imag'In   for supporting this work. }

\small
\vfill\pagebreak


\bibliographystyle{IEEEbib}
\bibliography{maBiblio}

\end{document}